\documentclass[twocolumn,english,aps,superscriptaddress,floatfix]{revtex4-2}

\usepackage{amsthm}
\usepackage{amsmath}
\usepackage{graphicx}
\usepackage{amssymb}
\usepackage{bm}
\usepackage{latexsym}
\usepackage{braket}
\usepackage[arrowdel]{physics}
\usepackage{xcolor}
\usepackage{hyperref}
\makeatletter
\@ifundefined{textcolor}{}
{%
 \definecolor{BLACK}{gray}{0}
 \definecolor{WHITE}{gray}{1}
 \definecolor{RED}{rgb}{1,0,0}
 \definecolor{GREEN}{rgb}{0,1,0}
 \definecolor{BLUE}{rgb}{0,0,1}
 \definecolor{CYAN}{cmyk}{1,0,0,0}
 \definecolor{MAGENTA}{cmyk}{0,1,0,0}
 \definecolor{YELLOW}{cmyk}{0,0,1,0}
 }

\newcommand{\bE}{\bar{\mathcal{E}}}
\newcommand{\bB}{\bar{\mathcal{B}}}

\makeatother
\usepackage{babel}

\begin{document}

\author{J. Hainge}
\affiliation{Department of Physics and Astronomy, McMaster University, 1280 Main St.\ W., Hamilton, ON, L8S 4M1, Canada}
\author{N. Miladinovic}
\affiliation{Department of Physics and Astronomy, McMaster University, 1280 Main St.\ W., Hamilton, ON, L8S 4M1, Canada}
\author{D.\ H.\ J.\ O'Dell}
\affiliation{Department of Physics and Astronomy, McMaster University, 1280 Main St.\ W., Hamilton, ON, L8S 4M1, Canada}

\title{The Optical HMW Geometric Phase and the Abraham Minkowski Controversy}
\date{\today}

\begin{abstract}
\label{sec:abstract}
The question of the correct formulation for the momentum of light in a dielectric medium is typically referred to as the ``Abraham-Minkowski controversy".  Experiments conducted to elucidate the issue have primarily focused on measuring forces and momentum transfers.  In this work, we propose an interferometric approach using matter waves to measure the light-induced version of the He-McKellar-Wilkens (optical HMW) phase for a neutral atomic dipole in dynamical electromagnetic fields.  Beginning from the action principle, we show that this geometric phase is directly related via the Euler-Lagrange equations of motion to the Abraham force lying at the heart of the controversy.
\end{abstract}


\maketitle

The level of control that can be exerted over atoms using light lies behind some of the highest precision measurements made to date. This includes optical lattice atomic clocks with 18 digit or better precision that can measure sub-centimeter gravitational redshifts \cite{Oelker19,McGrew18,Bothwell22}, and  atom interferometers based on optical beam splitters, mirrors or lattices that enable the fine structure constant to be measured via atomic recoil to better than one part in a billion \cite{Parker18,Yu19,Morel20,Schelfhout24}, or  measure the gravitational acceleration due to miniature masses at the level of nm/s$^2$  \cite{Panda24, Overstreet22}. These levels of precision are not just useful for fundamental physics but are also relevant to high fidelity quantum technologies based on atoms and light \cite{Bongs19,Zhang24,Robicheaux21}. 

In this context it  is surprising that the momentum delivered by light to a neutral dielectric medium, which could be a dilute cloud of atoms or even a single atom, has been the subject of debate and apparently conflicting experimental results for over a century \cite{Pfeifer07, Baxter10, Penfield67, Brevik79, Milonni05, Anghinoni22}.  There are two main competing formulations for the momentum density of light in a medium: one due to Abraham, proposed in 1910: $\vec{g}_A \equiv \frac{1}{c^2}\vec{E}\cross\vec{H}$ \cite{Abraham09,Abraham10} (which is \emph{smaller} in a medium by a factor of the refractive index compared to the vacuum), and the other due to Minkowski, proposed in 1908: $\vec{g}_M \equiv \vec{D}\cross\vec{B}$ \cite{Minkowski10} (which is \emph{larger} in a medium by a factor of the refractive index compared to the vacuum) \cite{extensions}.  Here, $\vec{E}$, $\vec{D}$, $\vec{B}$, and $\vec{H}$ are the usual electric, displacement, magnetic and auxiliary magnetic field, respectively.  The differences between the two formulations are relativistic in origin \cite{Sonnleitner17}, and hence are usually small for cold atoms, but may have implications for future high precision measurements, especially those making use of large momentum transfers (LMT) or long (atom-light) interaction times \cite{Lan12, Chiow11, Kirsten23}.

Early experiments measuring the momentum transferred to mirrors suspended in dielectric fluids \cite{Jones51, Jones54, Jones78}, or the response of the surface of a fluid to radiation pressure \cite{Ashkin73} have given results consistent with $\vec{g}_M$, as have more recent, updated versions \cite{Verma15, Verma17, Capeloto16}, and a Bragg diffraction experiment using a Bose-Einstein condensate (BEC) \cite{Campbell05}.  However, other experiments have led to the opposite conclusion, including \cite{James68} and \cite{Walker75}, where the former measured forces induced in a dielectric by a time-varying electromagnetic field, and the latter measured the torque induced in a dielectric by a laser. Both of these experiments set out to measure the  Abraham force
\begin{equation}
	\vec{F}_A = \int \frac{\varepsilon_r \mu_r - 1}{c^2}\pdv{t}\vec{S} \dd V
	\label{EqAForce}
\end{equation}
a derivation of which may be found in \cite{Moller52} or \cite{Pauli81}. Here $c$ is the vacuum speed of light, $\vec{S} \equiv \vec{E}\cross\vec{H}$ is the Poynting vector, and $\mu_r \ (\varepsilon_r)$ is the relative permeability (permittivity) of the medium.  The refractive index is $n = \sqrt{\varepsilon_r\mu_r}$.  In the simplest case of an isotropic and non-dispersive medium, Eq. (\ref{EqAForce}) is the only difference in the equations of motion which follow from Abraham and Minkowski's respective formulations of the stress-energy tensor for light in a medium.  The detection of this ``extra" mechanical force supports Abraham's formulation, while its absence supports Minkowski's.  More recent experiments focused on the detection of this force include  \cite{Brevik12, Rikken12, Zhang15, Kundu17, Choi17}.

\begin{figure}[!h]
\includegraphics[width=0.825\columnwidth]{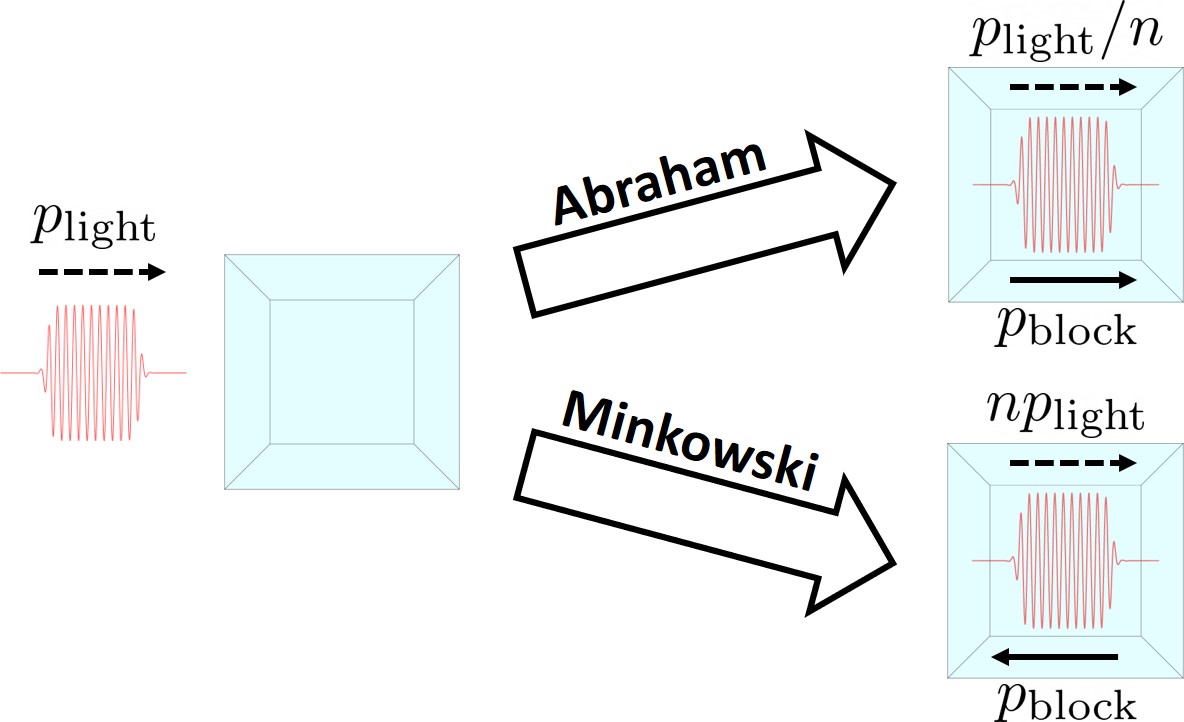}
	\caption{Balazs thought experiment \cite{Balazs53}. A light pulse enters a block of glass which recoils (anti)parallel to the light under the (Minkowski) Abraham formulation.  In this work, we follow \cite{Hinds09} and replace the glass with a single atom or BEC.}
	\label{fig:Balazs}
\end{figure}

A thought experiment due to Balazs illustrates the difference between $\vec{g}_{A}$ and $\vec{g}_{M}$ \cite{Balazs53,Barnett10b}. A pulse of light is fired at a block of dielectric as shown in Fig.\ \ref{fig:Balazs}; arguments based on conservation of momentum and center of mass energy predict that $\vec{g}_{A}$ ($\vec{g}_{M}$) will cause the block to be pushed to the right (left). We emphasize that the total momentum is conserved in both cases. In  \cite{Hinds09} the slab is replaced by a single atom which allows first principles microscopic calculations to be performed that avoid the complications of a medium---a major factor in the controversy---while still retaining all the essential features of light interacting with an electrically polarizable object.  These calculations reveal that the standard gradient or dipole force $\vec{F}_{d}= d_i \grad E^i$ due to the leading edge of the pulse pulls the atom to the left (assuming red detuning), where $\vec{d}$ is the induced  dipole moment and the Einstein summation convention has been used. However, $\vec{F}_A$, which is commonly ignored, produces a force of twice the magnitude and to the right, giving an overall result in agreement with $\vec{g}_{A}$. Unfortunately, \cite{Hinds09} also concluded that random kicks due to spontaneous emission would mask the difference in displacement between $\vec{g}_{A}$ and $\vec{g}_{M}$.

\begin{figure}[t]
\includegraphics[width = 0.75\columnwidth]{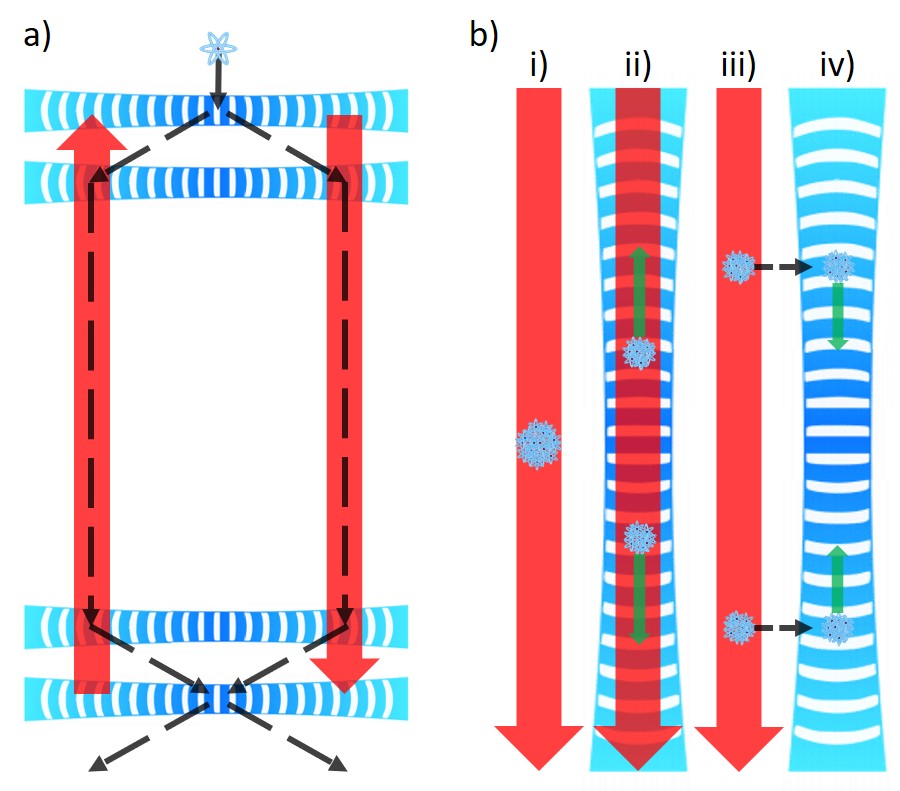}
	\caption{Two suggested setups for atom interferometers  to measure the OHMW phase while cancelling the kinetic and AC Stark phases. \textbf{a)} A Mach-Zehnder interferometer for atoms (black dashed). The four standing wave lasers (blue) act as beam-splitters and mirrors via Bragg scattering. A retro-reflected travelling wave laser (red) propagates in opposite directions along the two arms in order to generate the OHMW phase. \textbf{b)} Single beam implementation.  (i) A BEC already within the laser is (ii) Bragg scattered into a coherent superposition of moving with and against the laser Poynting vector.  (iii) The atom clouds are simultaneously ejected from the beam before being (iv) recombined  so they have equal interaction times with the laser.}
	\label{fig:MachZehnder}
\end{figure}

In this work, we break from the traditional approach to the Abraham-Minkowski controversy by avoiding any attempt at a measurement of forces or momentum transfers, and instead focus our attention at the level of the action $S = \int L\hspace{0.2em} \dd t$, which is, in our opinion, more fundamental.  The difference between the two formulations manifests itself as a Berry phase \cite{Berry84} in an atom interferometry experiment like the ones suggested in Fig. \ref{fig:MachZehnder}.  We will show that this phase is a light-induced (optical) version of the He-McKellar-Wilkens (HMW) phase acquired by a neutral electric dipole moving in a magnetic field.  The HMW phase was independently proposed by He and McKellar \cite{He93} and Wilkens \cite{Wilkens94HMW}, and has been measured using atom interferometry by the Toulouse group \cite{Lepoutre12,Gillot13,Lepoutre13a,Lepoutre13b}, although only for the case of \emph{static} electric and magnetic fields. It is the electromagnetic dual of the Aharonov-Casher phase \cite{Aharonov84, Dowling99, Horsley07}, and can be decomposed into the Aharonov-Bohm phases of the individual charges making up the dipole \cite{Wei95}. Hence, the optical HMW (OHMW) phase is sensitive to geometric and topological features of matter-light interactions. 

An advantage of an interferometric approach is that the phase continues to accumulate for as long as the atom interacts with the light even when no forces act, and remains once outside its influence. This is in contrast to measuring momentum or displacement, since momentum is imparted to the atom during the leading edge of a light pulse and is removed by the trailing edge, and there are no forces in between.  Moreover, the geometric nature of the phase confers the benefit of velocity-independence. Since the  OHMW phase is directly connected to the Abraham force via the Euler-Lagrange equations of motion, then, like the presence or absence of the Abraham force, the presence or absence of the phase is the essence of the controversy \cite{Hinds09, Leonhardt06PRA}. 

Why is $\vec{F}_{A}$ often ignored in cold atom experiments? Assuming linear response $\vec{d}=\alpha \vec{E}$, where $\alpha$ is the polarizability, and using the relation $n^2 = 1 + \frac{\alpha}{\varepsilon V}$ which holds for a dilute gas  \cite{MilonniBook05, volume}, where  $V$ is the volume and $\varepsilon$ the permittivity, one finds $F_{A}/F_{d}=(2/c) (\partial_{t} E^2)/(\partial_{x} E^2) \sim (2/c) \Delta x / \Delta t$. If the relevant velocity $v=\Delta x/\Delta t$ is that of an atom then $F_{A}/F_{d} \sim 2 v/c$ which is clearly small. However, $F_{A}=2F_{d}$ if the relevant velocity is that of a light pulse \cite{Hinds09}.  While $\vec{F}_d$ is present all the time the lasers are on, $\vec{F}_A$ is only non-zero when the atom experiences a time-varying Poynting vector.  This means that $\vec{F}_A$ is absent in experiments which use counterpropagating beams (for which $\vec{S} = 0$), for example, the experiment in \cite{Campbell05} which measured the recoil momentum of atoms within a BEC via a Kapitza-Dirac interferometer.  The measurement demonstrated that the atoms recoiled with a momentum directly proportional to the refractive index $n$, in agreement with Minkowski's formulation.  However, because the experiment used counterpropagating lasers, the Abraham and Minkowski formulations yield the \textit{same} momentum transfer ($\vec{F}_A$ is the only difference between the two).

The controversy can in fact be resolved in many ways because it is the (arbitrary) splitting of the total stress energy tensor of the system into a material and an electromagnetic component \cite{Ginzburg79, Mikura76} which distinguishes the two formulations.  The ``correct" formalism is then a matter of convenience, which may be dictated by the response of the medium \cite{Leonhardt14}. For example, the presence of a medium implies an absorbed/emitted photon need not match the recoil momentum of an absorbing/emitting particle, which is always given by $\vec{g}_{M}$, because the medium can account for any difference \cite{Mansuripur10, Milonni10, Bradshaw10}. An elegant resolution \cite{Barnett10} is that $\vec{g}_{M}$ and $\vec{g}_{A}$ lead to the canonical and kinetic momenta, respectively, and thus Minkowski's momentum is associated with wave properties (diffraction), while Abraham's is associated with particle properties (center of mass motion) \cite{Leonhardt06Nat}, in line with the idea that the controversy can be resolved by carefully distinguishing (particle) momentum from (wave) pseudomomentum \cite{Gordon73, McIntyre81, Peierls91}. 

Following \cite{Hinds09}, we choose our medium to be a single atom (or a BEC of non-interacting atoms) and for simplicity we treat only the non-dispersive case.  The effects of dispersion will be presented  elsewhere \cite{future}. Since we need to include relativistic effects we start from the relativistically invariant Lagrangian, for which the interaction term reads \cite{He17}
\begin{equation}
	L_{\mathrm{int}} = \frac{1}{2\gamma}\left(\vec{d}\cdot\vec{E} + \vec{m}\cdot\vec{B}\right)  \, ,
	\label{Lint}
\end{equation}
where $\gamma \equiv \sqrt{1 - \beta^2}^{-1}$ is the Lorentz factor for a particle moving with velocity $\vec{v} = \vec{\beta}c$, and $\vec{m}$ is the magnetic dipole moment of the atom.  Specializing to non-magnetic transitions and neglecting the out of phase component of the electric dipole moment $\vec{d}_0$, then in the rest-frame of the atom (where $\vec{m}_0 = 0$) we have
\begin{equation}
	L_{\mathrm{int},0} = \frac{1}{2}\vec{d}_0\cdot\vec{E}_0 = \frac{1}{2}\alpha_0 E_0^2 \, .
	\label{Lint0}
\end{equation}
The 0 subscripts indicate the rest frame (lab-frame quantities bear no subscript).  The rest frame interaction Lagrangian is then simply the negative of the AC Stark shift.  As we have in mind alkali atoms in their ground state interacting with linearly polarized light, we need not worry about vector or tensor contributions to the polarizability \cite{LeKien13, Bhowmik20}.  Boosting to the lab-frame electromagnetic fields yields 
\begin{equation}
	E_0^2 = \gamma^2\left[\left(\vec{E} + \vec{v}\cross\vec{B}\right)^2 - \left(\vec{E}\cdot\vec{\beta}\right)^2\right] \, ,
	\label{EEq}
\end{equation}
and substituting this and Eq.\  (\ref{Lint0}) into (\ref{Lint}) gives the exact lab-frame interaction Lagrangian
\begin{equation}
	L_{\mathrm{int}} = \frac{1}{2}\gamma\alpha\left[\left(\vec{E} + \vec{v}\cross\vec{B}\right)^2 - \left(\vec{E}\cdot\vec{\beta}\right)^2\right] \, .
	\label{LintFin}
\end{equation}

The $\vec{v}\cross\vec{B}$ term can be recognized as the motional electric field which gives rise to the R\"{o}ntgen interaction whereby a neutral but moving electric dipole interacts with a magnetic field  \cite{Sonnleitner17,Rontgen88}.
Note that a factor of $\gamma$ is ``lost" in taking $\dd t_0 \rightarrow \gamma^{-1}\dd t$ at the level of the action, and that by neglecting dispersion, we implicitly assume that $\alpha = \alpha_0$.  For a two level atom, the polarizability is \cite{Cohen-Tannoudji92}
\begin{equation}
	\alpha = -\frac{\abs{d_{ge}}^2}{\hbar}\frac{\delta}{\delta^2 + \frac{1}{4}\Gamma^2 + \frac{1}{2}\Omega^2} \approx -\frac{\abs{d_{ge}}^2}{\hbar\delta}
	\label{alpha}
\end{equation}
where $d_{ge}$ is the dipole matrix element between the ground and excited state, $\delta \equiv \omega_L - \omega_a$ is the detuning of the lab-frame laser frequency $\omega_L$ from the atomic transition frequency $\omega_a$, $\Gamma$ is the linewidth of the transition, and $\Omega \equiv -\hbar^{-1}\vert d_{ge} \vec{E} \vert$ is the Rabi frequency.  In this work, we assume that $\abs{\delta} \gg \Gamma, \Omega$, and that we are sufficiently red-detuned so as to be in a non-dispersive region of $\alpha$.  In particular, this means that the Doppler shift of $\alpha$ is neglected.

To obtain the equations of motion to leading (zeroth) order in $\beta$, we need only keep terms up to first order in the Lagrangian. Second order terms will contribute to the Euler-Lagrange equations either at first order in $\beta$, or proportionally to acceleration.  The latter gives an effective mass correction of the order $\frac{\alpha E^2}{mc^2}$, which, for atomic masses and reasonable laser intensities, is safely ignorable \cite{Sonnleitner18}.  For instance, taking the $D_2$ transition of $^{7}$Li \cite{Gehm03,Fischer23}, and assuming a 50 W Gaussian laser with a 100 $\mu$m waist and a wavelength of 10.6 $\mu$m (corresponding to a CO$_2$ laser), yields $\frac{\alpha E^2}{mc^2} \approx 1 \times 10^{-17}$, where the electric field is on the order of $10^6$ N/C, and $\alpha \approx 5 \times 10^{-39}$ C$^2$m/N. Note that with these  parameters, we would not be in a completely non-dispersive regime.  However, this affects the results of the phase calculation below only in a quantitative manner; the order of magnitude of the measurement remains unaffected.

Including the free particle Lagrangian $-\frac{mc^2}{\gamma}$ \cite{Becker64v2} in the action, we obtain at leading order the Euler-Lagrange equation 
\begin{equation}
	m\vec{a} = \alpha E_i \grad E^i + \pdv{t}\left(\vec{d}\cross\vec{B}\right) = \frac{\alpha}{2}\grad E^2 + \alpha\mu\pdv{t}\vec{S}
\label{EqEoM}
\end{equation}
where $\mu$ is the magnetic permeability of the atom.  The forces present are the dipole and R\"ontgen forces \cite{Baxter93, Wilkens94Rontgen, Sonnleitner17}. In fact, comparison of the last term in Eq.\ (\ref{EqEoM}) with Eq.\ (\ref{EqAForce}) using the relation $n^2  = 1 + \frac{\alpha}{\varepsilon V}$  reveals the R\"ontgen and Abraham forces to be one and the same in this context.

Consider now the interferometric approach. In the experiments that measured the static HMW phase  \cite{Lepoutre12,Gillot13,Lepoutre13a,Lepoutre13b}, a supersonic ($v\approx 1000$ m/s) beam of lithium atoms was passed through a Mach-Zehnder interferometer where each arm traversed a different side of a double capacitor. The electric fields in the capacitor polarized the atoms in the two arms in opposite directions,  and together with a common magnetic field generated an HMW phase 
\begin{equation}
	\phi_{\mathrm{HMW}} = \hbar^{-1} \oint \left(\vec{B}\cross\vec{d}\ \right)\cdot\dd \vec{r} \, .
	\label{HMW}
\end{equation}
A key part of the experiment was the cancellation of dynamical phases due to kinetic energy, Stark and Zeeman shifts along each arm which together gave rise to a phase in excess of $300$ rad, leaving only the geometric phase $\phi_{\mathrm{HMW}} = 27 $ mrad. As shown in Fig.\ \ref{fig:MachZehnder}(a), this set-up can be adapted to measure the OHMW phase by replacing the static fields by electromagnetic fields in two laser beams travelling in opposite directions along the arms. We assume that the entrance of the atoms into the lasers has negligible contribution to the accumulated phase, because the gradient force may only change the transverse velocity of the atom, which has no impact on the phase accumulated, and the change in velocity along the beam direction for the proposed experimental parameters is on the order of $10^{-9}$ m/s.  Further, we assume the amount of time needed to reach the center of the beam is small compared to the time the atom will spend travelling inside the beam.   Once near the center of the beam, both forces in Eq.\ (\ref{EqEoM}) become negligible but their potentials continue to generate phase.  Integrating the Lagrangian in Eq. \ref{LintFin} within this context, we obtain a phase which is directly related to the Abraham force experienced by the atom upon entering the laser, and hence the momentum of the light while ``inside" the atom.  

Taking $\vec{E} = \bE\cos{\omega_L t}$, we time-average over an optical cycle to compute the phase $\phi$ generated along the right arm of the interferometer in Fig.\ \ref{fig:MachZehnder}.  To first order in $\beta$
\begin{equation}
	\hbar\phi_{R} \approx \int^T_0 \left(\frac{1}{2}mv^2 + \frac{\alpha}{4}\left[\bE^2 - 2\vec{\beta}\cdot\left(\vec{\bE}\cross\vec{\bB}c\right)\right] \right)\dd t
	\label{topphase}
\end{equation}
where $\frac{1}{2}\bE^2 \equiv \frac{\omega_L}{2\pi}\int_0^{2\pi/\omega_L} \vec{E}\cdot\vec{E} dt$, and $\vec{\bE}\cross\vec{\bB}$ is similarly defined.  The integration is over the interaction time, $T$, between the atom and the light.  Along the left arm, the only difference is the reversal of the relative directions of the laser and the atomic velocity.  Hence the phase difference between the two paths is given by
\begin{equation}
	\Delta\phi = \phi_{R} - \phi_{L} \approx -\int^T_0 \frac{\alpha}{\hbar}\left(\vec{\bE}\cross\vec{\bB}c\right)\cdot\vec{\beta} \dd t
	\label{intphase}
\end{equation}

The AC Stark and kinetic phases, being even in $\beta$, are identical along each arm, and so cancel out.  If we define $\vec{d}^* \equiv \alpha\vec{\bE}$ \cite{dstar}, then we can rewrite the integral to elucidate its geometric nature 
\begin{equation}
	\begin{aligned}
		\Delta\phi &\approx \hbar^{-1}\int^T_0 \left(\vec{\bB}\cross\alpha\vec{\bE}\right)\cdot\vec{v} \dd t = \frac{1}{2\hbar}\oint \left(\vec{\bB}\cross\vec{d}^*\right)\cdot\dd\vec{r} \\
		& \equiv \phi_{\mathrm{OHMW}} 
	\end{aligned}
	\label{phase}
\end{equation}
which defines the OHMW phase. We see that written in terms of $\vec{d}^*$ and $\vec{\bB}$, the OHMW phase takes half the value of the static HMW phase. Like all geometric phases, it does not depend upon velocity, only on the path taken through the fields, thereby making it robust against velocity dispersion in the atomic beam. To estimate its magnitude we assume that the atoms and light co-propagate over a length $L$ and also that to leading order $\sqrt{\bE^2}=c\sqrt{\bB^2}$, so that 
\begin{equation}
	 \phi_{\mathrm{OHMW}} \approx -\frac{\alpha\bE^2}{\hbar c}L  \, .
\end{equation}
Putting $L = 5$ cm, and using the previously stated values for $\alpha$ and laser intensity gives $\phi_{\mathrm{OHMW}} = -20$ mrad.  Increasing the laser power or interaction length are the most obvious changes which would boost the signal. 

One might worry about the effects of spontaneous emission on the visibility of the signal.  We note here that the saturation parameter \cite{Cohen-Tannoudji92} for the chosen experimental configuration is $s \equiv \frac{\frac{1}{2}\Omega^2}{\delta^2 + \frac{1}{4}\Gamma^2} \approx 10^{-8}$, such that the probability of occupying the excited state is $p_2 \equiv \frac{1}{2}s/(1 + s) \approx 10^{-8}$.  Hence we expect roughly a time of $\frac{1}{\Gamma p_2} \approx 2$s before a spontaneous decay event, where $\Gamma \approx 6$ MHz for the D$_2$ transition of $^{7}$Li \cite{Gehm03,Fischer23}, a time which is long compared to the atom-laser interaction time.  We therefore neglect any loss in contrast we might accrue due to spontaneous emission events.

The main difficulty in measuring the OHMW phase will be in separating it from the phase due to the AC Stark shift of the ground state. The latter is a dynamical phase and so depends on the atom-light interaction time, but putting $T=L/v$ its magnitude is
\begin{equation}
\phi_{S} \approx \frac{1}{2}\frac{\alpha\bE^2}{\hbar v } L  \, .
\end{equation}
Thus, the ratio is $\phi_{\mathrm{OHMW}}/\phi_{S}=2v/c \approx 10^{-5}$ if we assume $v=1000$ m/s. Any uncertainty in $\phi_{S}$ can therefore easily overshadow $\phi_{\mathrm{OHMW}}$ (the corresponding ratio for the successful experiment on the static case was $10^{-4}$).  This presents a formidable challenge to any experiment since the two counterpropagating laser beams cannot be expected to have exactly the same intensity, nor the trajectories of the two atom beams be exact mirror images of each other. However, we note that $\phi_{S}$ can be distinguished from $\phi_{\mathrm{OHMW}}$ because the former depends on interaction time, whereas the latter does not, and thus varying the velocity of atoms would discriminate between the two. This allows some possibility for detecting imbalance and accounting for it.

An alternative scheme is depicted in Fig.\ \ref{fig:MachZehnder}(b), for which the underlying physics remains essentially identical.  By placing a non-interacting BEC in a single laser beam (ideally with a super-gaussian or flat-top profile \cite{Gillen16}), and then coherently splitting it via an LMT beamsplitter, it would be possible to mitigate the issues with the two-laser scheme.  The current best LMT schemes generate velocities of the order of a few hundred atomic recoils \cite{Chiow11,Wilkason22}.  While this results in a ratio $\phi_{\mathrm{OHMW}}/\phi_{S} \approx 10^{-7}$ two orders of magnitude lower than the previous scheme, here we expect the Stark phase can be cancelled to much higher accuracy; the use of a single laser ensures that both atom clouds see the same laser intensity, while the flat-top beam profile lessens the impact of any misalignment between the lasers and atomic trajectories (relative to the case of a Gaussian beam). For example, using a second order super-Gaussian beam profile and assuming atomic velocities of 400 recoils, we expect that for atomic trajectories aligned with the laser axis to within 0.02 of a degree, and with a distance of closest approach to the center of the beam not farther than 0.02 of the beam width, the Stark phase would cancel out to the level of a few radians, while the OHMW phase remains largely unchanged.  This gains us an additional 4--5 orders of magnitude in the ratio between $\phi_{\textrm{OHMW}}$ and the leftover $\phi_{S}$.  One can also conceive of schemes that use light shift compensation, which has been demonstrated in high precision atom interferometry \cite{Kim20} to remove to $\phi_{S}$, but reverse the laser beam direction to preserve $\phi_{\textrm{OHMW}}$.

In this work, we have drawn a connection between the light-induced version of the He-McKellar-Wilkens geometric phase and the oft-neglected R\"ontgen (equivalently, Abraham) force, within the context of the Abraham-Minkowski controversy. As an alternative to previous measurements based on force, momentum, or displacement, we propose atom interferometry as a direct way to measure this phase, which we calculate beginning from the fundamental principle of a relativistically invariant action, as in \cite{He17}.  Although a measurement of the phase would be challenging with current capabilities, the increasing precision of measurements suggests that this phase will need to be accounted for in future experiments and technology based on atom-light interactions.

\acknowledgments 
Discussions with Xiao-Gang He, E. A. Hinds, Alan Jamison, Bruce McKellar, Guanchen Peng, Ben Sauer, Aephraim Steinberg,  and  Jacques Vigu\'e are gratefully acknowledged.

\bibliographystyle{apsrev4-2}

\begin{thebibliography}{90}
\bibitem [{vol()}]{volume}%
  \BibitemOpen
  \href@noop {} {\bibinfo {title} {{One might wonder about the appropriate
  volume to use, here, since the ``medium" is being treated as point-like.
  According to \cite{Hinds09}, the appropriate choice is the volume of the
  laser pulse which interacts with the atom. In this work, however, the
  refractive index itself is beside the point; the relevant parameter to
  consider is not the refractive index, $n$, but the polarizability, $\alpha$,
  which has no volume dependence.}}}\BibitemShut {Stop}%
\bibitem [{dst()}]{dstar}%
  \BibitemOpen
  \href@noop {} {\bibinfo {title} {{Note that $\vec{d}^* =
  \alpha\vec{\bar{\mathcal{E}}}$ is distinct from the relativistically correct
  definition of the lab frame electric dipole moment, which would instead be
  $\vec{d} = \alpha\left(\vec{\bar{\mathcal{E}}} + \vec{v}\times
  \vec{\bar{\mathcal{B}}}\right)$. The reason we use $\vec{d}^*$ in the main
  text, rather than $\vec{d}$, to draw our parallel with the HMW phase is to
  make explicit all of the velocity dependence in the Lagrangian, and to
  demonstrate that $\Delta\phi$ is truly a velocity-independent geometric
  phase.}}}\BibitemShut {Stop}%
\end{thebibliography}



%

\end{document}